\documentclass[sigconf]{acmart}
\AtBeginDocument{%
  }

\setcopyright{acmlicensed}
\copyrightyear{2025}
\acmYear{2025}
\acmDOI{10.1145/3746027.3755536}
\acmConference[MM '25]{Proceedings of the 33rd ACM International Conference on Multimedia}{October 27--31, 2025}{Dublin, Ireland}
\acmISBN{979-8-4007-2035-2/2025/10}

\acmSubmissionID{4317}

\usepackage{multirow}
\usepackage{microtype}
\begin{document}

\title{MoTAS: MoE-Guided Feature Selection from TTS-Augmented Speech for Enhanced Multimodal Alzheimer's Early Screening}


\author{Yongqi Shao}
\affiliation{%
\institution{Shanghai Jiao Tong University}
\city{Shanghai}
\country{China}}
\email{cici_syq@sjtu.edu.cn}

\author{Bingxin Mei}
\affiliation{%
\institution{Shanghai Jiao Tong University}
\city{Shanghai}
\country{China}}
\email{mei18895349540@sjtu.edu.cn}

\author{Cong Tan}
\affiliation{%
\institution{Shanghai Jiao Tong University}
\city{Shanghai}
\country{China}}
\email{coign@sjtu.edu.cn}

\author{Hong Huo}
\affiliation{%
\institution{Shanghai Jiao Tong University}
\city{Shanghai}
\country{China}}
\email{huohong@sjtu.edu.cn}

\author{Tao Fang}
\affiliation{%
\institution{Shanghai Jiao Tong University}
\city{Shanghai}
\country{China}}
\email{tfang@sjtu.edu.cn}



\begin{abstract}
Early screening for Alzheimer's Disease (AD) through speech presents a promising non-invasive approach. However, challenges such as limited data and the lack of fine-grained, adaptive feature selection often hinder performance. To address these issues, we propose MoTAS, a robust framework designed to enhance AD screening efficiency. MoTAS leverages Text-to-Speech (TTS) augmentation to increase data volume and employs a Mixture of Experts (MoE) mechanism to improve multimodal feature selection, jointly enhancing model generalization. The process begins with automatic speech recognition (ASR) to obtain accurate transcriptions. TTS is then used to synthesize speech that enriches the dataset. After extracting acoustic and text embeddings, the MoE mechanism dynamically selects the most informative features, optimizing feature fusion for improved classification. Evaluated on the ADReSSo dataset, MoTAS achieves a leading accuracy of 85.71\%, outperforming existing baselines. Ablation studies further validate the individual contributions of TTS augmentation and MoE in boosting classification performance. These findings highlight the practical value of MoTAS in real-world AD screening scenarios, particularly in data-limited settings.
\end{abstract}


\ccsdesc[500]{Applied computing~Health informatics}

\keywords{Alzheimer’s Disease (AD), Speech-Based AD Screening, Text-to-Speech (TTS) Augmentation, Mixture of Experts (MoE)}


\begin{teaserfigure}
  \includegraphics[width=\textwidth]{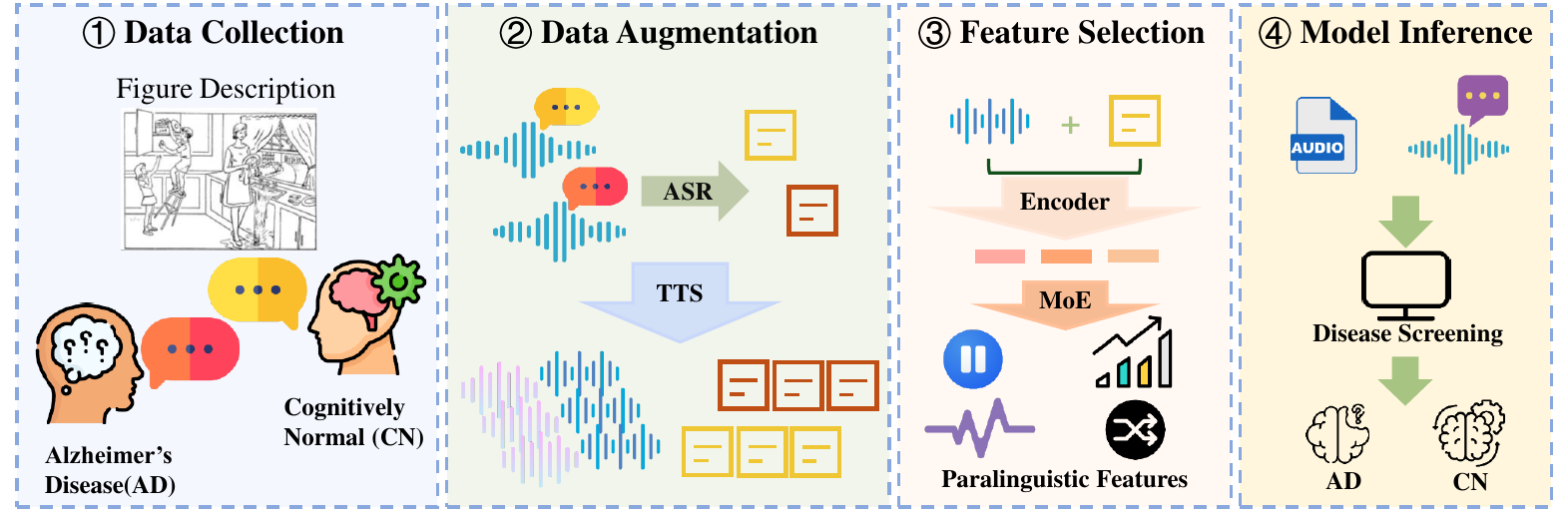}
  \caption{The Overall Pipeline of MoTAS: MoE-Guided Feature Selection from TTS-Augmented Speech for Multimodal Alzheimer’s Early Screening. Some Icons Adapted from Google Material Icons (\url{https://fonts.google.com/icons}).}
  \Description{The figure illustrates a four-stage pipeline for extracting and utilizing paralinguistic features in Alzheimer's disease detection. From left to right: (1) Data Collection – raw speech recordings are obtained from participants describing the “Cookie Theft” picture; (2) Data Augmentation – recordings are segmented into sentence-level utterances and augmented using TTS techniques to expand the dataset; (3) Feature Selection – key acoustic features such as pitch, speech rate, loudness, pauses, repetitions, and voice quality are extracted using specialized modules, each represented with intuitive icons; (4) Model Inference – speech inputs are fed into a pre-trained model to output classification results for cognitive impairment screening.}
  \label{fig:teaser}
\end{teaserfigure}

\maketitle

\section{Introduction}

Alzheimer’s disease is a progressive neurodegenerative disorder that primarily affects cognitive functions, memory, and language abilities. As the most common cause of dementia, its prevalence is rising sharply, with an estimated 55 million people affected globally. This number is projected to reach 139 million by 2050 due to aging populations\cite{jeon2024post}.

The growing burden of AD poses significant challenges to healthcare systems, leading to escalating care costs, economic strain, and profound social impacts. Despite extensive research into potential treatments, no cure currently exists. This underscores the critical importance of early diagnosis in slowing disease progression and improving patient outcomes.

Traditional methods for diagnosing AD include clinical assessments, neuroimaging techniques such as magnetic resonance imaging (MRI), positron emission tomography (PET) scans, and cerebrospinal fluid (CSF) biomarker analysis. While these methods provide valuable insights into disease progression, they are expensive, require specialized resources, and are unsuitable for large-scale early screening \cite{passeri2022alzheimer, jha2020alzheimer}. CSF testing also involves invasive lumbar puncture, causing discomfort and reducing compliance \cite{bharati2022dementia, mcgeer1986comparison}. Moreover, such tools often detect AD only at later stages, limiting the effectiveness of potential interventions. These limitations underscore the urgent need for a non-invasive, cost-effective, and scalable early detection approach.

Recent studies indicate that speech-based analysis is a promising alternative for AD detection, as language impairments often appear in the early stages. AD patients typically show speech features such as pauses, hesitations, reduced fluency, pronunciation errors, and lexical deficits \cite{yang2022deep}. With machine learning and deep learning models, speech analysis can automatically and objectively capture subtle linguistic and acoustic patterns linked to cognitive decline \cite{yang2022deep, luz2018method}. Compared to conventional diagnostics, it offers a more practical solution for early screening and timely intervention.

However, despite its potential, existing speech-based AD detection methods still face several challenges\cite{ding2024speech}. The limited availability of datasets constrains the generalization ability of deep learning models across diverse populations, making them prone to overfitting. Additionally, many models treat all features equally without adaptive selection, limiting their ability to capture fine-grained cues like speech rhythm and articulation errors. Moreover, state-of-the-art deep learning approaches often require substantial computational resources, limiting their practicality for real-time clinical applications.

To address these limitations, we propose MoTAS, a speech-based Alzheimer's screening framework that leverages TTS-augmented speech and MoE-guided feature selection. Figure~\ref{fig:teaser} illustrates the MoTAS pipeline. Our key contributions include:

\begin{itemize}
\item We propose a TTS data augmentation strategy that synthesizes speech associated with both AD and cognitively normal(CN) control groups, aiming to mitigate data scarcity and enhance model generalization.
\item A MoE-guided feature selection mechanism is introduced to adaptively select features from acoustic and linguistic modalities, thereby optimizing feature utilization and reducing redundancy.
\item Extensive experiments on the ADReSSo dataset demonstrate that our proposed framework significantly outperforms existing speech-based methods, achieving an accuracy of 85.71\%.
\end{itemize}

 By addressing challenges related to data availability, feature selection efficiency, and computational constraints, our approach advances automated, non-invasive, and scalable AD screening, providing a practical solution for early detection in real-world applications.


\begin{figure*}[h]
  \centering
  \includegraphics[width=\textwidth]{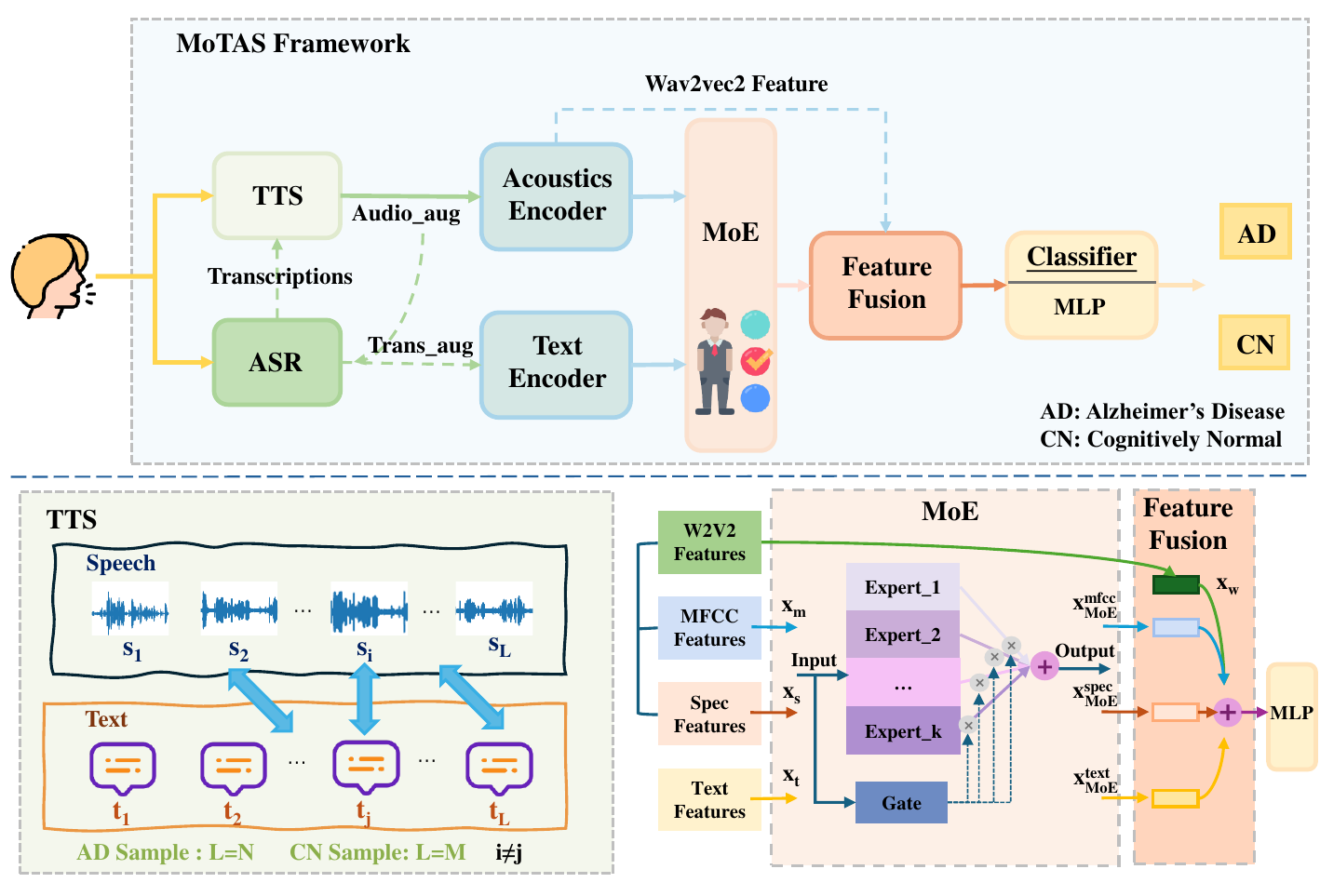}
  \caption{The Upper Part of the Diagram Illustrates the Framework of Our Proposed MoTAS; The Lower Part Details the Structure of the TTS Module on the Left and the MoE and Feature Fusion Modules on the Right.}
  \Description{This figure illustrates the overall framework of the proposed method, which integrates TTS-augmented speech and ASR-transcribed text for Alzheimer’s disease screening. Features are extracted using Wav2Vec2, ResNet18, BiLSTM, and BERT. A MoE mechanism dynamically selects the most informative features, which are fused and classified to distinguish between Alzhemier's Disease (AD) and cognitively normal (CN) subjects.}
  \label{fig:framework}
\end{figure*}

\section{Related Work}

 In this section, we first introduce the key advances in TTS technology for data augmentation, then discuss the role of MoE in adaptive feature selection, and finally review existing speech-based AD detection methods. 
 
\subsection{TTS for Data Augmentation in Speech Processing}

In recent years, TTS technology has made significant advancements, transitioning from concatenative and parametric models to deep learning-driven approaches. Modern TTS models, such as Fish-Speech\cite{liao2024fish}, VITS\cite{kim2021conditional}, WaveNet\cite{van2016wavenet}, and Tacotron\cite{wang2017tacotron}, have greatly enhanced the naturalness and intelligibility of synthesized speech. These models utilize sequence-to-sequence architectures and advanced waveform generation techniques, enabling the production of high-quality speech with human-like prosody and articulation. As a result, TTS has been widely adopted in assistive technologies, virtual assistants, and speech synthesis research.

TTS for data augmentation has proven effective in speech-related tasks, particularly in addressing data scarcity. It has been applied in ASR\cite{yang2025enhancing, do2024improving}, speech emotion recognition(SER)\cite{latif2023generative, praseetha2022speech}, and accent adaptation\cite{do2024improving, tan2021aispeech}, significantly improving model robustness and performance. However, its potential remains largely unexplored in AD detection. Current AD speech datasets are often small and imbalanced, making models prone to overfitting and limiting their generalizability. In this study, we leverage TTS-augmented synthetic AD speech to expand the dataset, preserving critical linguistic and acoustic features associated with cognitive decline. This enhancement boosts classification accuracy and improves the reliability of speech-based AD screening.

\subsection{MoE-Guided Adaptive Feature Selection}

Mixture of Experts (MoE) is a deep learning approach that improves model efficiency by dynamically selecting specialized subnetworks (experts) for different input types. It has been widely used in NLP, speech, and vision tasks to enhance scalability in large-scale learning. Recent models like Google’s Switch Transformer \cite{fedus2022switch} and DeepSeek-V3 \cite{liu2024deepseek} show that MoE can reduce computation while preserving high capacity.

In speech-related tasks, MoE has been employed to optimize feature selection and processing, leading to improved model generalization. Research has shown its effectiveness in ASR \cite{hsu2021hubert}, speaker verification\cite{wang2025mixture,gaur2021mixture}, and SER\cite{hyeon2024improving, liu2024moge, salman2025mixture},where it efficiently allocates different experts to process prosodic, phonetic, and spectral features, outperforming conventional deep learning approaches. Despite its success in various speech tasks, MoE has been underutilized in speech-based AD detection. While existing AD classification models often process acoustic and linguistic features separately, they typically lack mechanisms to adaptively prioritize the most informative cues within each modality, such as speech rhythm, articulation errors, and lexical patterns. Furthermore, many methods treat all input features equally, which may lead to suboptimal learning and inefficient computation.

To address these limitations, we propose an MoE-guided mechanism that adaptively selects expert networks based on feature types, enabling dynamic focus on salient linguistic and paralinguistic cues. This hierarchical design enhances AD detection accuracy while reducing redundancy and computational cost.

\subsection{Speech-Based AD Detection Methods}

Existing speech-based AD detection methods primarily extract acoustic and linguistic features from spontaneous speech recordings and their transcriptions\cite{mahajan2021acoustic, cui2021identifying, luz2021detecting}. Traditional approaches rely on handcrafted features, such as speech rate, pause duration, pitch variation, and Mel-Frequency Cepstral Coefficients (MFCCs), which are then classified using Support Vector Machines (SVMs) and Random Forests\cite{luz2018method, balagopalan2021comparing, chen2021automatic}. Additionally, linguistic features derived from text transcriptions, such as lexical diversity, syntactic complexity, and word repetition patterns, have been explored to detect early cognitive decline\cite{eyigoz2020linguistic, fraser2015linguistic}.

With the rise of deep learning, feature extraction has shifted from manual engineering to data-driven learning, significantly improving model performance. CNNs and RNNs have been successfully applied to Mel spectrograms and raw audio waveforms, capturing complex temporal and spectral variations\cite{gupta2021comparing}. Meanwhile, self-supervised learning models such as Wav2Vec2\cite{gauder2021alzheimer} enable feature extraction directly from raw speech, eliminating the need for manual feature engineering. In text analysis, Transformer-based models such as BERT\cite{devlin2019bert} and DistilBERT\cite{sanh2019distilbert} have been widely adopted to analyze transcribed speech, learning semantic coherence and syntactic changes associated with AD\cite{mirheidari2021identifying, pan2021using}. Furthermore, multimodal models integrating speech and text features have shown superior classification performance by leveraging complementary acoustic and linguistic markers\cite{wang2021modular, zhu2021wavbert, li2024whisper}.

Despite these advancements, existing methods still face critical challenges, including limited dataset availability, suboptimal feature fusion, and high computational costs. To address these issues, we propose MoTAS, a multimodal framework that increases dataset size using TTS and enhances feature selection via a MoE mechanism. By combining synthetic speech generation with adaptive, fine-grained multimodal features selection, our approach enhances the robustness, accuracy, and scalability of automated AD detection, making it more practical for real-world deployment.

\section{Methodology}

The overall framework of our method is illustrated in Figure~\ref{fig:framework}. Raw speech is transcribed using ASR and augmented with TTS. Both real and synthetic speech, along with their transcriptions, are encoded into multimodal features. These features are then refined through a MoE mechanism, subsequently fused, and used in the final step for classifying into AD or CN categories.

This section outlines the proposed MoTAS framework, concentrating on its key components: TTS for data augmentation, acoustic and text encoder, MoE for feature selection, and feature fusion and classification.

\subsection{TTS for Data Augmentation}

This study employs TTS to generate synthetic speech samples, thereby expanding the dataset while preserving disease-relevant acoustic characteristics. Since the original dataset consists solely of raw English speech recordings, we first apply ASR using Whisper \cite{radford2023robust} to obtain text transcriptions, creating paired audio-text data. The process is formulated as follows:

\begin{equation}
  T = f_{\text{ASR}}(S), \quad \text{where } f_{\text{ASR}} = \text{Whisper}
\end{equation}

Here, \( S \) and \( T \) represent the sets of raw speech samples and their corresponding text transcriptions, respectively. Let $S_{AD} = \{ s_1^{AD}, s_2^{AD}, \dots, s_N^{AD} \}$ and $S_{\text{CN}} = \{s_1^{\text{CN}}, s_2^{\text{CN}}, \dots, s_M^{\text{CN}}\}$ represent the sets of AD and CN speech samples, respectively; $T_{\text{AD}} = \{t_1^{\text{AD}}, t_2^{\text{AD}}, \dots, t_N^{\text{AD}}\}$ and $T_{\text{CN}} = \{t_1^{\text{CN}}, t_2^{\text{CN}}, \dots, t_M^{\text{CN}}\}$ denote the corresponding ASR-transcribed texts for AD and CN speech samples, where $N$ and $M$ are the number of AD and CN samples in the original dataset.

Once transcriptions are obtained, a pre-trained TTS model \( f_{\text{TTS}} \) is used to synthesize new speech samples. The generated synthetic speech retains the acoustic characteristics of the reference speaker while replacing the linguistic content with transcriptions from another speaker. The synthesis process is defined as:

\begin{equation}
  \hat{s}^{\text{AD}}_i = f_{\text{TTS}}(t^{\text{AD}}_j, s^{\text{AD}}_i), \quad
\hat{s}^{\text{CN}}_i = f_{\text{TTS}}(t^{\text{CN}}_j, s^{\text{CN}}_i)
\end{equation}

where \( i = 1, \dots, N \), \( j = 1, \dots, N \), and \( i \ne j \) for AD samples, and similarly \( i = 1, \dots, M \), \( j = 1, \dots, M \), and \( i \ne j \) for CN samples. Here, \( s_i \) provides the speaker identity and \( t_j \) provides the linguistic content. The TTS model synthesizes speech that combines the voice characteristics of \( s_i \) with the transcript \( t_j \).

To augment the dataset further, each speech sample \( s_i \) can be paired with multiple transcriptions \( t_k \) from the same class (where \( k \ne i \) and \( k \ne j \)). This intra-class pairing allows a single reference voice to be combined with various linguistic contents, producing a rich set of synthetic samples with consistent speaker identity and class label. Such augmentation improves data diversity while preserving class-specific characteristics.

In this study, we employ FishSpeech\cite{liao2024fish}, a state-of-the-art TTS model designed for high-fidelity speaker-preserving synthesis, ensuring that the generated speech retains the original speaker’s prosody, rhythm, and articulation. We also balance the proportion of real and synthetic samples during training to prevent overfitting. 

 Although the speech and text come from different speakers, the TTS model preserves key acoustic features specific to the original speaker, which is crucial for ensuring that the synthetic speech accurately reflects disease-related vocal cues. This approach allows the synthetic speech to remain authentic, accurately reflecting the acoustic traits of the original speaker, thereby enhancing the reliability and accuracy of the dataset.

The final augmented speech is defined as:

\begin{equation}
 S_{\text{aug}} = S_{\text{orig}} \cup \{\hat s_i^{\text{AD}}, \hat s_i^{\text{CN}}\}
\end{equation}

To prevent semantic redundancy, we rely on the fact that even when reusing textual content across speakers, the acoustic expression remains speaker-dependent due to variations in prosody and articulation. As a result, the ASR-transcribed text from real participants' speech naturally reflect speaker-specific disfluencies or omissions, introducing lexical variation that enhances diversity at both the acoustic and linguistic levels.

Thus, after generating the augmented speech samples, we perform a second round of ASR on the synthetic audio, with the resulting transcriptions serving as new textual inputs for subsequent processing. The augmented transcription process is defined as follows:

\begin{equation}
  T_{\text{aug}} = f_{\text{ASR}}(S_{\text{aug}}), \quad \text{where } f_{\text{ASR}} = \text{Whisper}
\end{equation}

Finally, the augmented dataset is denoted as:

\begin{equation}
  \text{Data}_{\text{aug}} = \left\{ S_{\text{aug}},\ T_{\text{aug}} \right\}
\end{equation}

This paired multimodal dataset, together with the original dataset, serves as the input for downstream tasks including feature extraction, selection, fusion, and classification in our framework.

Overall, the TTS-augmented speech mechanism enhances dataset diversity while maintaining data quality, allowing the model to generalize more effectively across different speech patterns and mitigating overfitting issues. 

\subsection{Acoustic and Text Encoder}

For each speech sample $s_i$ and its corresponding text transcription $t_i$, we extract features from both acoustic and text modalities, forming the input feature set:

\begin{equation}
 X = \{x_w, x_m, x_s, x_t\} = \{ f_{\text{W2V2}}(s_i), f_{\text{MFCC}}(s_i), f_{\text{Spec}}(s_i), f_{\text{Text}}(t_i) \}
\end{equation}

where $f_{\text{W2V2}}(s_i)$ represents deep phonetic features extracted using Wav2Vec2\cite{baevski2020wav2vec}, capturing nuanced acoustic patterns that reflect speech prosody, phonetics, and articulation dynamics; $f_{\text{MFCC}}(s_i)$ denotes MFCC-based temporal dynamics modeled by BiLSTM\cite{huang2015bidirectional}, reflecting the spectral envelope and short-term temporal structure; $f_{\text{Spec}}(s_i)$ represents spectrogram-based features extracted using ResNet18\cite{he2016deep}, capturing the the local time-frequency energy distribution and prosodic cues; and $f_{\text{Text}}(t_i)$ corresponds to semantic and syntactic embeddings obtained from BERT\cite{devlin2019bert}, encoding the high-level semantic and syntactic patterns of speech. 

These features comprehensively represent both the acoustic and linguistic aspects of the data, allowing MoE to selectively integrate the most relevant multimodal information.

\subsection{MoE for Feature Selection}

Building upon Section 3.2, we introduce the MoE mechanism to dynamically select the most informative multimodal features, optimizing classification performance. For this mechanism, we consider the following subset of features without the Wav2Vec2 component:
\begin{equation}
 X_{\text{MoE}} = \{x_m, x_s, x_t\} = \{f_{\text{MFCC}}(s_i), f_{\text{Spec}}(s_i), f_{\text{Text}}(t_i)\}
\end{equation}

The exclusion of \( f_{\text{W2V2}}(s_i) \) is because this feature is extracted by a pre-trained model and already contains rich representational capabilities. Since it is primarily focused on representing acoustic features, further selection is not necessary. On the other hand, the text features extracted by BERT still contain valuable semantic information and are considered crucial for the task, thus they are retained in the feature selection phase. This distinction ensures that MoE can focus on integrating and enhancing the discrimination power of spectral, temporal, and semantic features.

The MoE mechanism consists of \(k\) expert networks, each designed to capture different patterns from the input feature vector. As shown in Figure~\ref{fig:framework}, the three types of features are all input into the same MoE mechanism, but the MoE process for each feature is performed independently. Each expert network in the MoE mechanism produces an output for its corresponding feature type, with \( x_m \), \( x_s \), and \( x_t \) being input separately as follows:

\begin{equation}
y_i = E_i(x), \quad i \in \{1, 2, \dots, k\}
\end{equation}

where \(E_i(\cdot)\) represents the expert network corresponding to each features.

To dynamically control the contribution of each expert, a gating network \( G(\cdot) \) is employed to generate a weight vector \( w \in \mathbb{R}^k \), where each element \( w_i \) corresponds to the importance of expert \( E_i \). For each feature type \( x \in \{x_m, x_s, x_t\} \), a separate gating network is used to compute the expert weights:

\begin{equation}
w = G(x) = \text{softmax}(W_g x + b_g)
\end{equation}

where \( W_g \in \mathbb{R}^{k \times d_x} \), \( b_g \in \mathbb{R}^{k} \), and \( d_x \in \{d_m, d_s, d_t\} \) denotes the dimension of the corresponding input feature. The softmax function ensures that the weights are positive and sum to 1, effectively determining the importance of each expert for the given input.

The final outputs of the MoE mechanism are computed separately for each feature type, providing distinct outputs for MFCC, spectrogram, and text features:
\begin{align}
x_{\text{MoE}}^{{mfcc}} &= \sum_{i=1}^{k} w_i^{\text{mfcc}} y_i^{\text{mfcc}}, \\
x_{\text{MoE}}^{{spec}} &= \sum_{i=1}^{k} w_i^{\text{spec}} y_i^{\text{spec}}, \\
x_{\text{MoE}}^{text} &= \sum_{i=1}^{k} w_i^{\text{text}} y_i^{\text{text}}
\end{align}
where \(y_i^{\text{mfcc}}, y_i^{\text{spec}}, y_i^{\text{text}}\) are outputs from the respective expert networks for MFCC, spectrogram, and text features, and \(w_i^{\text{mfcc}}, w_i^{\text{spec}}, w_i^{\text{text}}\) are the corresponding weights from the gating mechanism.

This independent processing ensures that the MoE mechanism effectively emphasizes the most relevant features for each type, enhancing the robustness and generalizability of the classification.

\subsection{Feature Fusion and Classification}

Following MoE-guided feature selection, we further enhance multimodal fusion by incorporating deep speech embeddings extracted via Wav2Vec2\cite{baevski2020wav2vec}. Unlike MFCC, spectrogram, and text features, which are compressed into unified representations through a modality-specific MoE mechanism, Wav2Vec2 embeddings are preserved in their raw or temporally-aggregated form. This approach leverages the pre-trained model's capacity for phonetic-level representation learning, thus avoiding unnecessary transformations and maintaining the integrity of low-level acoustic details.

To achieve comprehensive fusion of these diverse representations, we concatenate the MoE-guided features with the Wav2Vec2 feature:

\begin{equation}
x_{\text{final}} = \text{concat}(x_{\text{MoE}}^{{mfcc}}, x_{\text{MoE}}^{{spec}}, x_{\text{MoE}}^{{text}}, x_{\text{w}})
\end{equation}

where \( x_{\text{w}} \) represents the deep speech embeddings extracted by Wav2Vec2 in Section 3.2. The MoE-guided features provide high-level acoustic and linguistic information, while Wav2Vec2 captures phonetic and low-level speech characteristics. This dual-layered fusion strategy ensures that the final feature representation effectively integrates both high-level semantics and fine-grained acoustic details.

The multi-layer perceptron (MLP) classifier \( f_{\text{MLP}} \) consists of three fully connected layers with ReLU activations and dropout for regularization, tailored to handle complex interactions among fused features and prevent overfitting, particularly in data-limited clinical settingss:

\begin{equation}
f_{\text{MLP}}(x) = \text{FC}_3\left(\text{ReLU}\left(\text{Dropout}\left(\text{FC}_2\left(\text{ReLU}\left(\text{FC}_1(x)\right)\right)\right)\right)\right)
\end{equation}

The fused multimodal representation \( x_{\text{final}} \) is subsequently passed through the MLP classifier to produce a binary classification outcome:

\begin{equation}
y = f_{\text{MLP}}(x_{\text{final}})
\end{equation}

where \( y \in \{0,1\} \) denotes the classification results (AD vs. CN). The classifier is trained using a cross-entropy loss function to optimize accuracy and robustness:

\begin{equation}
L = - \sum_{i=1}^{N} \left[ y_i \log(\hat{y}_i) + (1 - y_i) \log(1 - \hat{y}_i) \right]
\end{equation}

Here, \( i \) represents the index of the \( i \)-th sample. By leveraging MoE for feature selection and Wav2Vec2 for deep speech embedding fusion, our approach achieves a balance between capturing discriminative multimodal information and phonetic details, thereby enhancing classification performance.

The method design separates the roles of adaptive multimodal selection via MoE and enhancement of low-level acoustic features through Wav2Vec2. MoE targets the most discriminative features across spectral, temporal, and semantic dimensions, while Wav2Vec2 enriches the model's capability to process phonetic irregularities and subtle speech characteristics.

These enhancements are aligned with clinical observations of Alzheimer's-related speech impairments, which include semantic disorganization and phonetic irregularities such as pauses and stuttering. The fusion of MoE-guided features with Wav2Vec2 phonetic embeddings provides a robust, hierarchy-aware representation, enhancing the detection model's expressiveness. We will demonstrate the effectiveness of this integrated framework with evaluations on the ADReSSo benchmark in subsequent sections.

\section{Experiments}

This section outlines the experimental setup used to assess the performance of the proposed framework. We first introduce the datasets and preprocessing steps, followed by the implementation details of model training and evaluation.

\subsection{Datasets and Data Preprocessing}

The dataset used in this study originates from the ADReSSo Challenge\cite{luz2021detecting}, which comprises English speech recordings of participants describing the “Cookie Theft” picture from the Boston Diagnostic Aphasia Exam\cite{goodglass2001boston}. Participants are categorized into two groups: CN and Probable AD. The original training set consists of 166 participants, while the test set includes 71 participants, with both sets balanced in terms of gender, age, and diagnostic category. Additionally, the recordings contain speech from experimenters providing instructions or engaging in brief conversations.

The original audio signals were sampled at 16kHz. During preprocessing, sentence-level timestamp annotations provided in the dataset were used to extract and segment the speech data. To ensure speaker consistency and semantic clarity, only the participant’s utterances were retained. For segments shorter than 5 seconds, the original content was preserved and zero-padded to meet the target duration. Silent or invalid segments, including those with ASR failures, were excluded to maintain data quality.

All transcriptions generated by Whisper ASR were further cleaned to improve consistency and alignment across samples. This cleaning process included converting all characters to lowercase, correcting spelling errors, and filtering out non-linguistic symbols. For each cleaned speech segment, both acoustic and textual features were extracted separately. The resulting segments preserve linguistic coherence while maintaining the original acoustic structure, providing high-quality inputs for subsequent multimodal analysis.

\begin{table}[t]
    \centering
    \caption{Dataset Splits Before and After 3$\times$ Augmentation}
    \label{tab:dataset}
    \begin{tabular}{lccc}
        \toprule
        & \textbf{Train} & \textbf{Train\_aug} & \textbf{Test} \\
        \midrule
        AD & 87  & 253  & 35  \\
        CN & 79  & 228  & 36  \\
        \midrule
        Total & 166 & 481 & 71  \\
        \bottomrule
    \end{tabular}
\end{table}

\subsection{Implementation Details}

To address the limited sample size of the ADReSSo training set, we employed a speaker-consistent TTS data augmentation strategy using the Fish-Speech toolkit. Specifically, for each subject, we synthesized new speech samples by reusing transcripts from other participants while preserving the original speaker’s acoustic characteristics. This approach maintains the speaker’s vocal identity while introducing semantic and lexical diversity. The augmented samples were generated proportionally to the original class distribution (AD vs. CN), thereby preserving label balance. Ultimately, the training set was expanded to approximately three times its original size. To evaluate the impact of different augmentation scales on model performance, we conducted comparative experiments using training sets expended by 1.5$\times$, 2$\times$, and 2.5$\times$. The optimal augmentation ratio was selected based on validation performance. A comparison of sample sizes before and after augmentation is shown in Table~\ref{tab:dataset}. In these ablation studies, each setting retained the original training data and supplemented the remaining portion with newly generated augmented samples as needed.

To capture acoustic characteristics at different levels, we extracted three types of acoustic features. MFCCs were computed using the Librosa library with 40 Mel filter banks, a 25\,ms frame length, and a 10\,ms hop size. The resulting MFCC sequences (13-dimensional per frame) were fed into a two-layer bidirectional LSTM (hidden size 128), and the final hidden state was passed through a fully connected layer to obtain fixed-length embeddings of dimension \(d_m = 128\). Mel spectrograms were resized to \(224 \times 224\) and processed by a pretrained ResNet18 with ImageNet weights. Segment-level features were aggregated using mean pooling, resulting in \(d_s = 1000\). Phoneme-level features were obtained by averaging the last hidden states of a wav2vec2-base-960h model, yielding \(d_{\text{w}} = 768\).

Textual features were derived from Whisper ASR transcripts. After preprocessing, each sentence was encoded using a pretrained BERT-base model, and the [CLS] token embedding was used as the sentence-level representation, with \(d_t = 1024\). All extracted features were stored for downstream alignment, fusion, and classification tasks.

For feature selection, we adopted a MoE mechanism, where each feature (MFCC, spectrogram, and text) was associated with three expert networks ($k=3$). A shared gating mechanism dynamically assigned weights to these experts based on the input. This framework enables the model to emphasize the most discriminative features and suppress redundant information, thereby improving performance and interpretability. Notably, Wav2Vec2 features were excluded from the MoE mechanism, as they already provide high-quality phonetic representations through self-supervised pretraining and were directly incorporated in the final fusion stage.

All model components were implemented using the PyTorch framework. Training was performed using the Adam optimizer with an initial learning rate of 0.0067 and a batch size of 32. Binary cross-entropy was used as the loss function. To ensure result reliability, each experiment was repeated five times with fixed random seeds, and the final performance metrics reported represent the average across all five runs.

\begin{table*}[h]
    \centering
    \caption{Comparison of Our Method With Existing Approaches on the ADReSSo Test Set. Metrics Include Accuracy, Precision, Recall, and F1-Score, Following Definitions From the Baseline Study \cite{luz2021detecting}. Our Method's Results Are Averaged Over Five Runs.}
    \label{tab:performance}
    \begin{tabular}{lcccccccc}
        \toprule
        \multirow{2}{*}{\textbf{Method}} & \multirow{2}{*}{\textbf{Accuracy (\%)}} & \multicolumn{2}{c}{\textbf{Precision (\%)}} & \multicolumn{2}{c}{\textbf{Recall (\%)}} & \multicolumn{2}{c}{\textbf{F1 Score (\%)}} \\
        \cmidrule(lr){3-4} \cmidrule(lr){5-6} \cmidrule(lr){7-8}
        & & AD & CN & AD & CN & AD & CN \\
        \midrule
        \multicolumn{8}{l}{\textbf{Without ASR Transcripts}} \\
        \midrule
        ADReSSo Baseline (eGeMAPS+SVM)\cite{luz2021detecting} & 64.79 & - & - & - & - & - & - \\
        Wav2Vec2+TB \cite{pan2021using}& 74.65 & 77.42 & 72.50 & 68.57 & 80.56 & 72.73 & 76.32 \\
        Whisper-TL medium\cite{li2024whisper} & 77.46 & 77.14 & 77.78 & 77.14 & 77.78 & 77.14 & 77.78 \\
        \midrule
        \midrule
        \multicolumn{8}{l}{\textbf{With ASR Transcripts}} \\
        \midrule
        ADReSSo Baseline (Late Fusion)\cite{luz2021detecting} & 78.87 & 77.78 & 80.00 & 80.00 & 77.78 & 78.87 & 78.87 \\
        WavBERT M$_b$ (W2V2 ASR + BERT)\cite{zhu2021wavbert} & 73.24 & 75.00 & 71.79 & 68.57 & 77.78 & 71.64 & 74.67 \\
        C-Attention-Unified\cite{wang2021modular} & 78.03 & 74.15 & 84.12 & 87.22 & 68.57 & 80.09 & 75.42 \\
        WavBERT M$_p$ (W2V2 ASR + BERT + Pauses)\cite{zhu2021wavbert} & 83.10 & \textbf{87.10} & 80.00 & 77.14 & \textbf{88.89} & 81.82 & 84.21 \\
        TDNN-ASR-M5\cite{pan2021using} & 84.51 & 81.58 & 87.88 & 88.57 & 80.56 & 84.93 & 84.06 \\
        Whisper-TL-FTP Medium\cite{li2024whisper} & 84.51 & 83.33 & 85.71 & 85.71 & 83.33 & 84.50 & \textbf{84.50} \\
        \textbf{MoTAS (Ours)} & \textbf{85.71} & 80.49 & \textbf{93.10} & \textbf{94.29} & 77.14 & \textbf{86.84} & 84.38 \\
        \bottomrule
    \end{tabular}
\end{table*}

\section{Results and Analysis}

This section presents the experimental results of our MoTAS framework for speech-based Alzheimer’s early screening, including comparisons with previous studies and an ablation study to evaluate the impact of key components.

\subsection{Comparison with Previous Studies}

We compared our proposed MoTAS framework with a range of existing speech-based AD detection models, including both acoustic-only approaches~\cite{luz2021detecting, pan2021using, li2024whisper} and multimodal methods that combine speech with ASR-transcribed text~\cite{luz2021detecting, zhu2021wavbert, wang2021modular, pan2021using, li2024whisper}. The comparative results are summarized in Table~\ref{tab:performance}.

As shown in the table, the proposed MoTAS framework achieves the highest overall classification accuracy (85.71\%) on the ADReSSo test set, outperforming all baselines from both single- and multi-modal categories. It also obtains the best CN precision (93.10\%), AD recall (94.29\%) and AD F1-score (86.84\%), indicating strong sensitivity and balanced detection performance. These results demonstrate the effectiveness of our design in capturing AD-related speech and language patterns with greater precision and robustness.

Compared to acoustic-only models such as Wav2Vec2+TB~\cite{pan2021using} and Whisper-TL medium~\cite{li2024whisper}, which achieve AD recall rates of 68.57\% and 77.14\% respectively, our framework shows substantial improvements. For example, MoTAS increases AD recall by over 17\% relative to Whisper-TL medium, while also improving accuracy and F1-score. These gains are likely attributed to the combined advantages of multimodal input, expert diversity, and MoE-guided adaptive feature selection, rather than data augmentation alone.

Among state-of-the-art multimodal systems, including WavBERT M$_p$~\cite{zhu2021wavbert}, TDNN-ASR-M5~\cite{pan2021using}, and Whisper-TL-FTP~\cite{li2024whisper}, our method remains the top-performing model. Although WavBERT M$_p$ achieves a strong AD precision of 87.10\% and CN recall of 88.89\%, MoTAS outperforms it across several key metrics, including AD F1-score (86.84\% vs. 81.82\%), AD recall (94.29\% vs. 77.14\%), and accuracy (85.71\% vs. 83.10\%). These results reflect the complementary benefits of TTS data augmentation and adaptive expert selection enabled by the MoE mechanism.

Notably, several multimodal baselines exhibit class imbalance. For example, WavBERT M$_b$~\cite{zhu2021wavbert} achieves only 68.57\% AD recall, while C-Attention-Unified~\cite{wang2021modular} shows a strong bias toward AD classification, achieving a recall of 87.22\% for AD and 68.57\% for CN. These outcomes suggest that naive modality fusion without adaptive control can lead to redundancy or modal dominance. In contrast, the MoE gating mechanism in our framework selectively emphasizes the most informative features for each input, improving both classification balance and model interpretability.

MoTAS also demonstrates robustness to ASR errors, which are common in spontaneous and cognitively impaired speech. The MoE gating mechanism effectively down-weights unreliable textual features, mitigating their impact on final predictions. Importantly, the high AD recall (94.29\%) and F1-score (86.84\%) are especially valuable in clinical screening scenarios, where reducing false negatives is critical for early diagnosis and intervention. By maintaining high sensitivity without compromising precision or overall accuracy, our model helps mitigate underdiagnosis risks.  

In summary, the proposed MoTAS framework combines multimodal inputs, TTS-augmented speech, and MoE-guided adaptive feature selection to achieve balanced and superior performance across both AD and CN classes, demonstrating strong potential for real-world deployment in early-stage AD screening based on spontaneous speech.

\begin{table*}[h]
    \centering
    \caption{Ablation Study on TTS Augmentation and MoE. We Evaluated Dataset Expansion at 1.5$\times$, 2$\times$, 2.5$\times$, and 3$\times$, Followed by MoE Ablation on Both the Original and Best-Performing Augmented Datasets. Results Are Averaged Over Five Runs on the ADReSSo Test Set\cite{luz2021detecting}.}
    \label{tab:tts_moe_performance}
    \begin{tabular}{c c c c c c c c c c c}
        \toprule
        \multirow{2}{*}{\textbf{ID}} & \multirow{2}{*}{\textbf{TTS (times)}} & \multirow{2}{*}{\textbf{MoE}} & \multirow{2}{*}{\textbf{Accuracy (\%)}} & \multicolumn{2}{c}{\textbf{Precision (\%)}} & \multicolumn{2}{c}{\textbf{Recall (\%)}} & \multicolumn{2}{c}{\textbf{F1 Score (\%)}} \\
        \cmidrule(lr){5-6} \cmidrule(lr){7-8} \cmidrule(lr){9-10}
        & & & & AD & CN & AD & CN & AD & CN \\
        \midrule
        1 & X & X & 78.28 & 81.46 & 76.06 & 73.71 & 82.86 & 77.20 & 79.20 \\
        2 & X & \checkmark & 79.71 & \textbf{82.58} & 77.98 & 76.00 & \textbf{83.43} & 78.81 & 80.40 \\
        3 & \checkmark(2) & X & 81.72 & 78.30 & 86.75 & 88.00 & 75.43 & 82.74 & 80.48 \\
        4 & \checkmark(2) & \checkmark & \textbf{85.71} & 80.49 & \textbf{93.10} & \textbf{94.29} & 77.14 & \textbf{86.84} & \textbf{84.38} \\
        5 & \checkmark(1.5) & \checkmark & 81.72 & 80.76 & 82.99 & 83.43 & 80.00 & 82.02 & 81.38 \\
        6 & \checkmark(2.5) & \checkmark & 82.86 & 79.29 & 87.73 & 89.14 & 76.57 & 83.87 & 81.68 \\
        7 & \checkmark(3) & \checkmark & 80.29 & 80.65 & 82.40 & 81.72 & 78.86 & 80.43 & 79.75 \\
        \bottomrule
    \end{tabular}
\end{table*}

\subsection{Ablation Study}

To evaluate the independent and synergistic contributions of the TTS-augmented speech data and MoE-guided feature selection mechanism in our framework, we conducted a comprehensive ablation study. As shown in Table~\ref{tab:tts_moe_performance}, the MoE mechanism was evaluated under two conditions: without data augmentation (Experiment ID 1 and 2) and with 2$\times$ TTS augmentation, which yielded the best performance (Experiment ID 3 and 4). The TTS augmentation was further analyzed by comparing multiple augmentation factors, including none, 1.5$\times$, 2$\times$, 2.5$\times$, and 3$\times$ (Experiment ID 2, 5, 4, 6, and 7, respectively).

\begin{figure}[h]
  \centering
  \includegraphics[width=\linewidth]{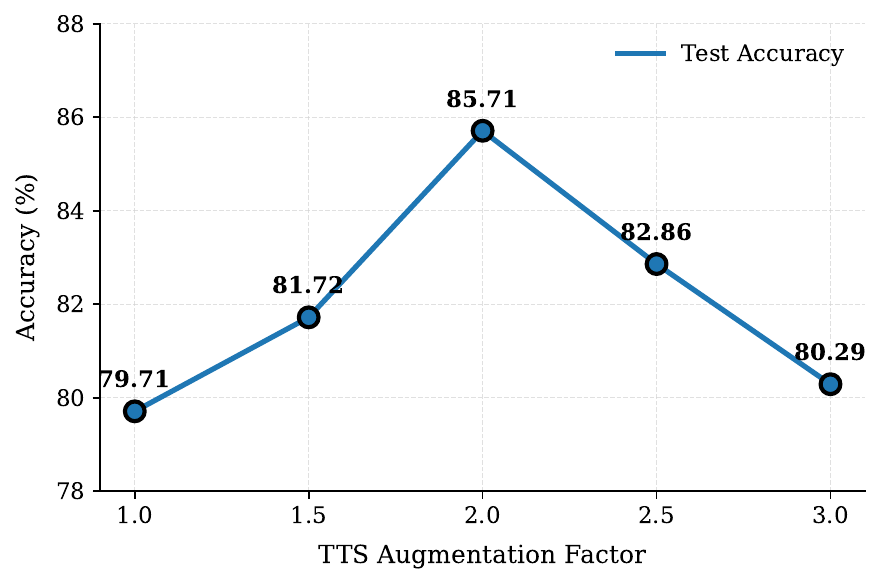}
  \caption{Test Accuracy on the ADReSSo Dataset \cite{luz2021detecting} With Different TTS Augmentation Factors}
  \Description{This figure presents the test accuracy results on the ADReSSo dataset for different TTS augmentation factors (1$\times$, 1.5$\times$, 2$\times$, 2.5$\times$, and 3$\times$). The accuracy increases up to 2× augmentation but declines with further expansion, indicating an optimal augmentation level.}
  \label{fig:accuracy}
\end{figure}

The results demonstrate that the MoE mechanism consistently enhances performance across settings. Without TTS augmentation, introducing MoE increased the test accuracy from 78.28\% to 79.71\% (ID1 vs. ID2), indicating its effectiveness under limited data conditions. When applied to the 2$\times$ augmented dataset, MoE further improved accuracy from 81.72\% to 85.71\% (ID3 vs. ID4), representing a notable 3.99\% gain. In addition to accuracy, other metrics such as precision, recall, and F1-score also improved significantly, confirming MoE’s role in boosting robustness and discriminative capability. By dynamically weighting the importance of multimodal features, the MoE mechanism effectively reduces redundancy and enhances the model's ability to capture AD-relevant acoustic and linguistic characteristics.

For the TTS agumentation, Figure~\ref{fig:accuracy} illustrates the influence of varying augmentation levels on test accuracy. As the augmentation factor increased from none to 2$\times$, the accuracy steadily improved, peaking at 85.71\%. However, further augmentation to 2.5$\times$ and 3$\times$ led to a decline in accuracy to 82.86\% and 80.29\%, respectively. This performance degradation may be attributed to the reduced proportion of real samples, which leads the model to overfit to the synthetic distribution and impairs its generalization ability.

Therefore, effective data augmentation should not only focus on increasing quantity but also ensure the quality of synthetic data. Moderate augmentation improves sample diversity and mitigates overfitting, while excessive augmentation can negatively impact performance. Based on these findings, we selected 2$\times$ TTS augmentation as the optimal configuration, balancing dataset richness with training stability.

In summary, the ablation study validates the complementary strengths of our MoTAS framework. TTS enhances data diversity and generalization, while MoE improves the selection of discriminative features. The integration of both significantly boosts classification accuracy and robustness, providing a solid foundation for scalable and effective speech-based AD screening.

\section{Conclusion}

This study proposes an innovative framework MoTAS, which combines TTS-augmented speech data with MoE-guided feature selection to improve speech-based AD early screening. By expanding the training set with synthetic speech and adaptively selecting multimodal features, the proposed approach effectively addresses key challenges such as data scarcity, feature redundancy, and model overfitting.

Experiments on the ADReSSo dataset demonstrate that our method significantly outperforms existing speech-based models in both accuracy and robustness. The results confirm the synergistic effect of TTS augmentation and MoE-guided feature selection, which enhances model generalization while optimizing multimodal fusion under constrained computational resources.

This framework offers a flexible foundation for developing more efficient cognitive screening systems. Future work will explore its applicability to cross-lingual and cross-dataset scenarios, as well as further optimize computational efficiency to enable real-time clinical deployment. Overall, our findings highlight the potential of synthetic data generation and adaptive feature fusion in advancing early Alzheimer’s screening toward more efficient, reliable, and scalable solutions.

\bibliographystyle{ACM-Reference-Format}
\bibliography{syqdraft}










\end{document}